\documentstyle[prl,aps,epsf]{revtex} 
\begin{document}
\draft

\twocolumn[\hsize\textwidth\columnwidth\hsize\csname %
@twocolumnfalse\endcsname

\title
{Phase Separation of the Two-Dimensional $t-J$ model}

\author{ C.T. Shih$^1$, Y.C. Chen$^2$, and T.K. Lee$^3$}
\address{
$^{1}$Dept. of Physics, National Tsing Hua Univ., Hsinchu, Taiwan\\
$^{2}$Dept. of Physics, Tunghai Univ., Taichung, Taiwan\\
$^{3}$
Inst. of Physics, Academia Sinica, Nankang, Taipei, Taiwan
}

\date{\today}
\maketitle 
\begin{abstract} 
The boundary of phase separation of the two-dimensional $t-J$ model
is investigated by the power-Lanczos method and Maxwell construction.
The method is similar to a variational approach and it determines
the lower bound of the phase separation boundary with 
 $J_c/t=0.6\pm 0.1$ in the limit 
 $n_e\sim 1$. 
In the physical interesting regime 
of high T$_c$ superconductors where  $0.3<J/t<0.5$ 
 there is no phase separation. 

\end{abstract}
                                                                               
]

It is believed
that the main physical properties  of the high-temperature
superconductors 
can be described by the two-dimensional (2D) $t-J$
model on a square lattice. The Hamiltonian is:
\begin{equation}
H=-t\sum_{<i,j>\sigma} (\tilde{c}^+_{i\sigma}\tilde{c}_{j\sigma}+h.c.)
  + J\sum_{<i,j>}({\bf S}_i\cdot{\bf S}_j-{1\over4} n_in_j),
\label{e:tjm}
\end{equation}
where $<i,j>$ is the nearest-neighbor pairs and $\tilde{c}_{i\sigma}
=c_{i\sigma}(1-n_{i,-\sigma})$.
In this model the two terms compete with each other. The
kinetic term favors the phase which the electrons are homogeneously
distributed in the plane to minimize the kinetic energy. While the
exchange term attracts the electrons together to lower the magnetic
energy. It is easy to see that for very large $J/t$ the system will phase 
separate into a hole-rich region and a region without holes to maximize
the magnetic energy gain. 

There are experimental evidences as well as 
theoretical studies that indicate phase separation and
superconductivity are closely related. It is even
argued that the driving mechanism of superconductivity is the
same as that of phase separation \cite{dagotto94} or
superconductivity comes from the frustrated phase separation
\cite{emery93}. Hence it is extremely important to determine the phase
separation boundary of the 2D $t-J$ model to resolve these issues.
This paper reports our findings of the phase separation boundary.

Experimentally, phase separation of the superconducting
La$_2$CuO$_{4+\delta}$ compound are observed by several measurements
\cite{jorgensen88,hammel90,hammel91,chou96}.
The compound phase separates for $0.01\leq\delta\leq0.06$
below T$_{ps}\approx300K$ into the nearly stoichiometric
antiferromagnetic La$_2$CuO$_{4+\delta_1}$ with $\delta_1$ less
than 0.02 and N\'eel temperature $T_N\approx250K$, and a metallic
superconducting oxygen-rich phase La$_2$CuO$_{4+\delta_2}$
with $\delta_2\approx0.06$
with $T_c\approx34K$. The Sr doped compound
La$_{2-x}$Sr$_x$CuO$_{4+\delta}$ also phase separates for $x\leq0.03$
into superconducting
La$_{2-x}$Sr$_x$CuO$_{4+\delta'}$ ($\delta'\approx0.08$) and
nonsuperconducting La$_{2-x}$Sr$_x$CuO$_{4+\delta''}$
($\delta''\approx0.00$) phases
\cite{cho93}.
Recent muon spin resonance and nuclear quadrupole resonance
 experiments \cite{cho92,borsa94,borsa95}
on  La$_{2-x}$Sr$_x$CuO$_4$ also indicate 
 that the doped holes
were inhomogeneously distributed mesoscopically and segregated
into walls separating the hole-poor antiferromagnetic domains.

Theoretically, there are conflicting results. The first important 
paper on this issue is by 
Emery {\it et al.}\cite{emery90}. They used the exact
diagonalization (ED) to study the $4\times4$ cluster. Using 
Maxwell construction
they claimed that phase separation occurs for all values of $J/t$.
 This result is 
contradictory to the  later calculations by using
quantum Monte Carlo (QMC) \cite{moreo91} and
ED \cite{dagotto92} on the Hubbard model, which should be consistent
with the $t-J$ model for small $J/t$. 
Putikka {\it et al.} studied this problem using the
high-temperature series expansion and found phase
separation at T=0 for $J/t$ lying above a line extending from
$J/t=3.8$ at zero filling to $J/t=1.2$ at half filling
\cite{putikka92}. Prelov\v sek {\it et al.}\cite{prelovsek93}
calculated the two-point and four-point density correlations using
ED on clusters of size 18 and 20 sites. They found the two-hole bound state for
$J/t>0.2$. For $J/t>1.5$
the holes form domain walls along (1,0) or (0,1) direction,
 and phase separate into a hole-rich and a
hole-free phase for even larger $J/t>2.5$. 
Hellberg {\it et al.} determined very accurately
that the critical $J/t$ for phase separation at low
electron density limit is $J/t=3.4367$ \cite{hellberg95}.
 Poilblanc calculated the energy of 2 and 4
holes by ED on several clusters  up to 26 sites.
The phase diagram includes a liquid of d-wave hole pairs
for $J/t \geq 0.2$, a liquid of hole droplets (quartets)
for larger $J/t \geq 0.5$, and at even larger $J/t$, an
instability towards phase separation \cite{poilblanc95}.
 Yokoyama {\it et al.} investigated the phase diagram by
the variational  Monte Carlo (VMC) method \cite{yokoyama96}. The
critical $J/t$ for phase separation at the high density
limit they found is 1.5, which is consistent with Putikka
{\it et al.}. 

Most recently  Hellberg and Manousakis\cite{hellberg96} investigated
this problem by the Green Function Monte Carlo (GFMC)
method and Maxwell construction
 for larger clusters. Their phase diagram is similar to 
Emery
{\it et al.} \cite{emery90}. They conclude that the $t-J$ model phase separates
for all values of $J/t$ in the low doping regime.

The theoretical results of different groups discussed above
are consistent at the large $J/t$ and low electron density
region. But unfortunately, in the interesting physical regime
of high T$_c$ superconductors,  $0.3<J/t<0.5$ and high electron
density $0.75<n_e<0.95$, they are in disagreement. 
We have used the power-Lanczos 
 (PL) method \cite{yctk95,heeb93}
 to obtain the best estimate of the ground state
energy in this physical regime 
for the largest cluster (82 sites) that have been studied 
so far. Based on the variational argument we show
 that there is no phase separation
in this physical regime\cite{khono97}.

\begin{figure}[ht]
\epsfysize=9.cm\epsfbox{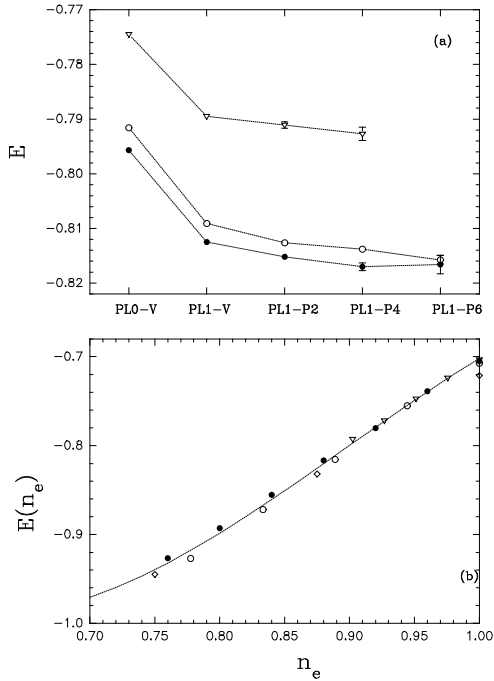}
\vskip 4mm
\caption{(a): Typical plots of energy per site vs powers for $J/t=0.6$,
$n_e=32/36$ (open circles) and $n_e=44/50$ (full circles) $n_e=74/82$ 
(open triangles); (b)Energy per site as a function of electronic
density for
$J/t=0.6$ with different cluster sizes.
Diamonds are the exact result of 16 sites. Open circles are for
36 sites and full circles are for 50 sites, both are obtained
by PL1$_{power=6}$. Triangles are for 82 sites with PL1$_{power=4}$.}
\label{f:psfsa}
\end{figure}

The ground state energy of the Hamiltonian of equation (1) is calculated
by using the PL method. The PL method we used is similar to the 
GFMC method but without using importance sampling and the fixed node
approximation. The method is essentially a variational approach. Applying
more powers to a trial wave function
implies a better approximation of the ground state wave function. 
Details of the method are discussed in  Ref.
\cite{yctk95}. 
The trial wave
functions we used are the  optimized  Gutzwiller wave functions,
resonating valence bond state (RVB) \cite{gros88}, and RVB with
antiferromagnetic long range order  \cite{tkct97}.
In Fig. 1(a) energy per site is plotted as a function of
power for $J/t=0.6$ and three different densities:
$n_e=32/36$ (open circles) and $n_e=44/50$ (full circles) $n_e=74/82$ 
(open triangles).  Error bar is 
shown only when it is larger than the symbol.
We also compared our energy of $J/t=1$ for 50/64 with the result
of high-temperature series expansion \cite{putikka97}.
The best energy we get is -1.183(2) while the high-temperature
expansion result is -1.20(2). They are well in agreement.
In Fig. 1(b) we show the best energies we are able to obtain for
clusters with 36, 50 and 82 sites as a function of electronic density.
For comparison we also show the exact energies of 16 sites\cite{dagotto92}.
Energies are little lower for the smaller clusters. For 50 and 82 sites,
there seems to be very little finite size effect. The energy per site
is a fairly smooth function of density. We do not find large effect due to
different Fermi surface topology in the physical regime.

To find the phase 
separation boundary  by using Maxwell construction 
we are interested in the variation of the slopes in
figures like Fig. 1(b). In other words we are interested in the second 
derivative of energy with respect to the electronic density, or the inverse
compressibility. It turns out that there is a systematic variation of this
quantity as the energy  approaches the ground state or as the power increases
in our PL method.  Although
in the physical regime
most of our best data have not yet converged to the exact ground state, this
systematic variation is enough for us to determine the 
lower bound of the phase separation boundary.

It is difficult to read out the slope variation from
figures like Fig.1(b), as the curve is almost a straight line for
$n_e>0.85$. Therefore we shall follow  
 Emery {\it et al.}\cite{emery90} by examining another quantity.
In the one-dimensional $t-J$ model  the phase separated
state contains an electron-free and a electron-rich phases. However, 
it phase separates into a hole-free
phase, i.e., the antiferromagnetic Heisenberg island, and a
hole-rich phase 
in
the two-dimensional $t-J$ model. 
Thus the energy of the phase separated state is in the form:
\begin{equation}
E=(N_s-N)e_H+Ne_h
\label{e:ps1}
\end{equation}
where $N_s$ is the total number of sites and N is the number of sites
in the hole-rich phase. $e_H=1.169J$ denotes the Heisenberg energy
per site \cite{manousakis91}. And $e_h$ is energy
per site in the uniform hole-rich phase, which is a function of
the hole density in this phase $x=N_h/N$. $N_h$ is the
number of holes. E can be rearranged into the form:
\begin{equation}
E=N_se_H+N_he(x)
\label{e:ps2}
\end{equation}
where
\begin{equation}
e(x)\equiv[-e_H+e_h(x)]/x
\label{e:ps3}
\end{equation}
If e(x) of a particular $J/t$ has
a minimum at $x=x_m$ and the hole density of the total system is
smaller than $x_m$, the system will adjust the size of the
hole-rich phase N such that $x_m$ is equal to $N_h/N$ and it minimizes the total
energy in Eq.(\ref{e:ps2}). Since $N_s$, $e_H$, and $N_h$ are all
constants, the total energy is minimized as $e(x)$ is minimized. Thus
$x_m$ is the critical density for phase separation at this $J/t$.

We calculated $e(x)$ from the energy of the uniform states $e_h(x)$
by the PL method and found the minimum of $e(x)$ on
$6\times6$, $\sqrt{50}\times\sqrt{50}$, and $\sqrt{82}\times\sqrt{82}$
clusters for several densities and $J/t$.
It is very difficult to get the converged ground state
energy in the physical regime due to the
sign problem. 
After we have found the optimized wave function in the 
VMC calculation we used the PL
method to project the trial wave function onto the ground state
systematically. The PL-1 power=4 (for 82 sites) or PL-1 power=6
(for 50 and 36 sites) energy
is used here as the $e_h(x)$. It is about $2\sim4$ percent lower than
the variational energy. We estimate the difference between the
best PL energy is within one or two percentage of the true ground state 
energy.

\begin{figure}[ht]
{\hspace*{1mm}
\epsfysize=13cm\epsfbox{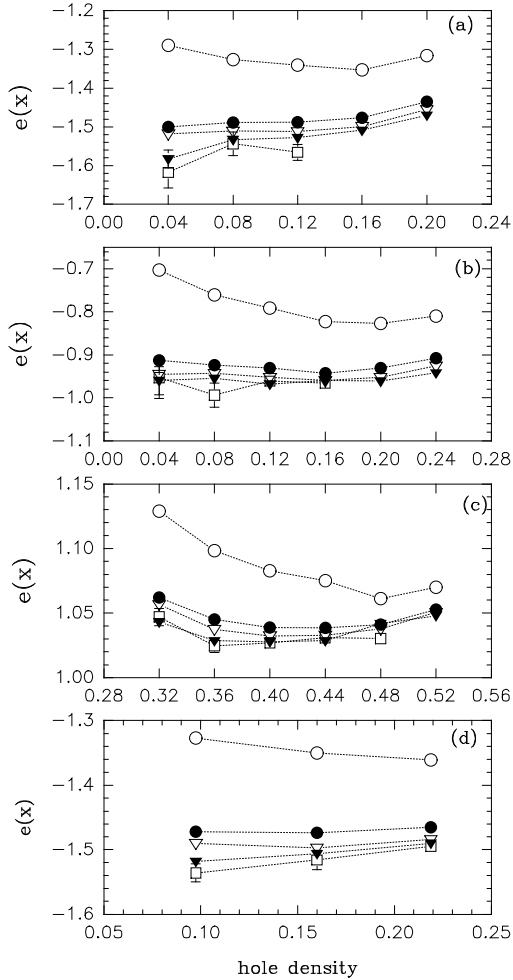}
}
\vskip 4mm
\caption{$e(x)$ vs hole density $x$ for (a)$J/t=0.4$, (b)
$J/t=0.6$ and (c)$J/t=1.5$ for several powers: PL0-VMC (open
circles), PL1-VMC (full circles), PL1-power=2 (open triangles),
PL1-power=4 (full triangles), and PL1-power=6 (open squares).
(d)$J/t=0.4$ for close shells for different size of lattices,
74/82, 42/50, and 50/64.}
\label{f:pspwex}
\end{figure}

$e(x)$ vs $x=1-n_e$ calculated on 50 sites for $J/t$=0.4, 0.6
and 1.5 is shown in Fig.\ref{f:pspwex}(a)-\ref{f:pspwex}(c), respectively. 
It is interesting
to note the trend of the shift of $e(x)$ with powers. For
$J/t=0.4$ (Fig.\ref{f:pspwex} (a)), at the VMC level, the
minimum of $e(x)$ is at $x_m=0.16$. It shifts to $x=0.04$
(the minimum hole density we calculated for this cluster) immediately
after the first order Lanczos improvement (PL1-VMC) and stays
at the density up to 6 powers. For $J/t=0.6$ (Fig.\ref{f:pspwex}(b)),
$x_m$ shifts from $x=0.2$ (VMC) to $x=0.16$ (PL1-VMC) and to
$x=0.08$ (PL1-power=6) at last. For $J/t=1.5$ (Fig.\ref{f:pspwex} (c)),
$x_m$ shifts from $x=0.48$ (PL1-VMC) to $x=0.4$ (PL1-power=2) and to
 $x=0.36$ (PL1-power=6) at last. It is clear that $x_m$ shifts monotonically
 toward
 a smaller value  when the energy moves closer to the
ground state. 

The results presented 
 in Fig.\ref{f:pspwex}(a)-\ref{f:pspwex}(c) are calculated with a fixed lattice size and different electron numbers. Hence Fermi surfaces have
different shapes and, in particular, there are open and closed shells. 
It has been argued\cite{hellberg96} that comparing energies obtained 
for these different Fermi surfaces might be inaccurate. To examine this
argument carefully, we have compared systems with closed shell Fermi
surfaces only. 
In Fig.\ref{f:pspwex}(d)  $e(x)$ calculated
from close shells of different size of lattices for $J/t=0.4$ shows similar
behavior as  Fig.\ref{f:pspwex}(a). The minimum of $e(x)$
shifts toward smaller hole density. The trend of $x_m$ moving
with increasing power is the same for both close and open shells.
Hence the shell effect is not important here.

\begin{figure}[b]
\epsfysize=6cm\epsfbox{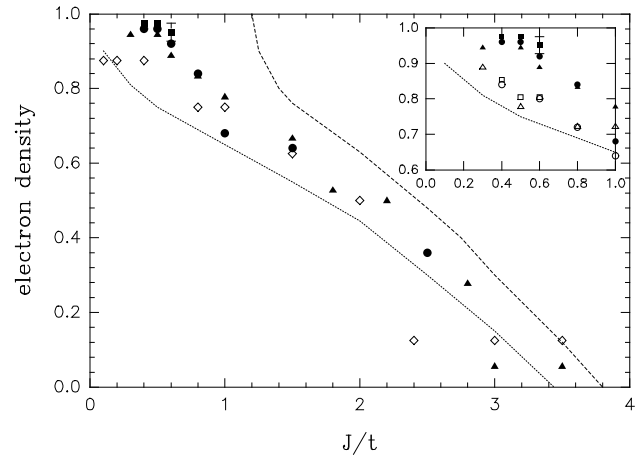}
\vskip 4mm
\caption{Phase separation boundary on the phase diagram of the
two-dimensional $t-J$ model evaluated by ED on the $4\times 4$
lattice[11] (open diamonds),
by the high-temperature series expansion[14] (dashed line)
by the GFMC  method[19] (dotted line), and the
PL method on 36 sites (full triangles), 50 sites (full
circles), and 82 sites(full square).
The  phase boundary determined by the VMC method for 36 sites (open
triangles) and 50 sites (open circles) are
shown in the inset.}
\label{f:PSboundary}
\end{figure}

In  Fig.\ref{f:PSboundary} we show the phase separation boundary determined
by the best $x_m$. 
The PL1-power=6 phase boundaries of 36 sites 
and 50 sites are shown as full triangles and full circles,
respectively.
Also some of the PL1-power=4 data of 82 sites are also
shown as full squares. For $J/t=0.6$ the error bars of the $e(x)$
for $n_e=$ 80/82, 78/82, and 76/82 are larger than the difference of these
three $e(x)$, thus error bars of $x_m$ are shown in the figure near
 these electron densities. 

The dashed line in Fig.\ref{f:PSboundary} is the result of 
high temperature series expansion\cite{putikka92}.  
Similar result is obtained by
the variational study 
\cite{yokoyama96}.
They assumed the system
separates into a hole-free Heisenberg antiferromagnet and an
electron-free vacuum state. This overestimates the energy required for
 the phase-separated state, 
 since electrons can "evaporate" from
the Heisenberg island to gain energy. 
 Their critical $J_c/t\approx1.2$ is larger
than our $J_c/t\approx0.6$.
Similar argument was also given by Hellberg  and Manousakis
\cite{hellberg96}.

Our estimate of the $J_c/t=0.6\pm 0.1$ is actually a lower bound. The exact
phase
separation boundary should be  to the 
right of our result in Fig.\ref{f:PSboundary}. When we use much poorer
estimate of the ground state energy as our VMC result, the phase
boundary is shifted lower. This is 
shown in the inset of Fig.\ref{f:PSboundary}. The VMC
results of 36 sites (open triangles) and 50 sites (open circles) show a
much smaller $J_c/t$. 

\unitlength 3.2mm
\begin{picture}(21,17)
\put(0.3,10){$x$}
\put(10.5,0){$J/t$}
\put(2.4,1.1){0.0}
\put(7.4,1.1){0.5}
\put(12.4,1.1){1.0}
\put(17.4,1.1){1.5}
\put(1.4,4.3){0.1}
\put(1.4,6.8){0.2}
\put(1.4,9.3){0.3}
\put(1.4,11.8){0.4}
\put(1.4,14.3){0.5}
\put(1.4,16.8){0.6}
\put(3,2){\begin{picture}(15.5,15)
\put(0,0){\line(0,1){15}}
\put(15.5,0){\line(0,1){15}}
\multiput(0,0)(0,2.5){6}{\line(1,0){0.4}}
\multiput(0,0.5)(0,0.5){30}{\line(1,0){0.2}}
\put(0,0){\line(1,0){15.5}}
\put(0,15){\line(1,0){15.5}}
\multiput(0,0)(1,0){16}{\line(0,1){0.2}}
\multiput(0.,0)(5,0){4}{\line(0,1){0.4}}
\put(4,1){\circle*{0.57271}}
\put(4,2){\circle*{0.45607}}
\put(4,3){\circle*{0.47434}}
\put(4,4){\circle*{0.39497}}
\put(4,5){\circle*{0.39243}}
\put(5,1){\circle*{0.55408}}
\put(5,2){\circle*{0.50596}}
\put(5,3){\circle*{0.42661}}
\put(5,4){\circle*{0.41713}}
\put(6,1){\circle*{0.50000}}
\put(6,2){\circle*{0.48270}}
\put(6,3){\circle*{0.41713}}
\put(6,4){\circle*{0.37815}}
\put(6,5){\circle*{0.36606}}
\put(6,6){\circle*{0.36332}}
\put(8,3){\circle*{0.44385}}
\put(8,4){\circle*{0.42895}}
\put(8,5){\circle*{0.36742}}
\put(8,6){\circle*{0.32000}}
\put(10,5){\circle*{0.36194}}
\put(10,6){\circle*{0.32879}}
\put(10,7){\circle*{0.30364}}
\put(10,8){\circle*{0.29950}}
\put(10,9){\circle*{0.25298}}
\put(15,8){\circle*{0.28618}}
\put(15,9){\circle*{0.27129}}
\put(15,10){\circle*{0.23622}}
\put(15,11){\circle*{0.21000}}
\put(15,12){\circle*{0.18193}}
\put(15,13){\circle*{0.14866}}
\put(2,12.6){\circle*{0.70711}}
\put(2.7,12.3){$\Delta e(x)=0.5$}
\end{picture}
}
\end{picture}
\begin{figure}
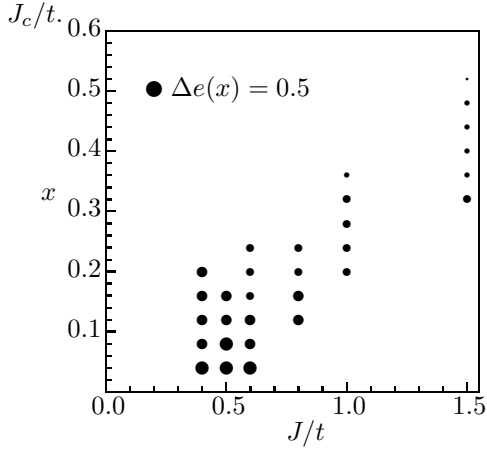

\caption{$e(x)$ difference
between PL1-power=6 and PL0-VMC for 50 sites. The values are proportional
to the area of the circles.}
\label{f:PSpercent}
\end{figure}

Another way to understand this argument of lower bound 
is to examine the variation of $e(x)$ with power.
In Fig.\ref{f:PSpercent} we show the change of $e(x)$
between PL1-power=6 and VMC for 50 sites. The values are proportional to 
the area of the circles. Because of the $1/x$ factor in Eq. (4), 
the smaller the hole density the
more improvement of $e(x)$ will likely occur. 
Because of the variational nature of the PL method, the
larger the improvement observed between VMC and PL1-power=6  the larger
the difference between the exact result and PL1-power=6 will be.
Hence, once the minimum $x_m$ is at the lowest
hole density such as $J/t=0.4$ in Fig.\ref{f:pspwex}, 
better estimate of the ground state energy by
applying more powers will not change the minimum to higher hole density.
Based on this argument we are confident to conclude that
there is no phase separation in the physical regime where $0.3<J/t<0.5$.

We have found that
for $J/t\leq0.5$, the minimum of $e(x)$ is always at two holes
for clusters of different sizes (16, 36, 50 and 82). 
As argued
by Dagotto {\it et al.},
this might indicate
 a two-hole bound state\cite{dagotto92} but not phase separation. 
 If there were phase separation, the $x_m$ would be
at the same (or nearby) density  rather than the
same number of holes. 

It is also interesting to note that in Fig.\ref{f:pspwex}(b), for $J/t=0.6$ 
the minimum $x_m$ seems to be
at 4 holes instead of 2 holes.  This is observed for both 36 and 50 sites.
It seems to be quite consistent with a recent
claim by Poilblanc\cite{poilblanc95} that there is a phase with  quartets
for 
$0.5\leq J/t\leq 0.8$. But our
data is not accurate enough for 82 sites to make a more definite conclusion.


 Recently Hellberg and Manousakis\cite{hellberg96}
have used GFMC to determine the phase separation boundary.
 The phase boundary they reported
(dotted line in Fig.\ref{f:PSboundary}) is similar to our variational
boundary (see the inset of Fig.\ref{f:PSboundary}). 
Without knowing details of their calculation
we cannot completely understand  this discrepancy. 
A possible clue is that  they might not have obtained lower enough energy
in the high electron density regime. As shown in Fig.\ref{f:pspwex}, 
in particular Fig.\ref{f:pspwex}(d), until 
the energy is lower enough to be closer to the ground state, it is very easy to
to make the conclusion that there is a minimum of $e(x)$ at a finite hole
density.

In summary, we determined the phase separation boundary by the PL
method and Maxwell construction. We have studied various size of clusters
and densities of holes. The largest cluster studied is 2 holes in an 82
site lattice. 
Using the variational nature of the PL method and the systematic
variation of the energy as a function of hole density we
conclude that the 
 critical $J_c/t$ for phase separation in
the low hole density limit is at least $\approx0.6$.
There is no phase separation in the physical regime.

It should be pointed out that
the result reported above  are obtained by assuming  the
hole-rich region in the phase separated state has a uniform hole density.
We have not yet considered more exotic possibilities such as the 
stripe phase 
\cite{prelovsek93,tsunetsugu95,white96}.

We wish to thank C.S. Hellberg, H.Q. Lin and W.O. Putikka 
for many useful discussions. 
This work is partially supported by the National Science Council of
Republic of China, Grant Nos. NSC 86-2811-M-007-001R \&
86-2112-M-001-042T \& 86-2112-M-029-001.
Part of computations were performed at the National Center
for High-Performance Computing in Taiwan. We are grateful for
their support.

\end{document}